\documentclass[10pt,
aps,pra,
reprint,
notitlepage,
superscriptaddress,
frontmatterverbose, 
showpacs,
preprintnumbers,nofootinbib,
amsmath,amssymb,
]{revtex4-1}
\usepackage[combine]{ucs}
\usepackage[version=3]{mhchem} 
\usepackage{graphicx}
\usepackage[british]{babel}
\usepackage{ifpdf}
\usepackage{natbib}
\usepackage{multirow}
\usepackage{appendix}
\usepackage{hyperref}

\hypersetup{pdfauthor={Salvador Rodríguez Gómez},pdftitle={Controlling thermal expansion: A Metal--Organic Frameworks route}}
\begin{document}
\bibliographystyle{unsrtnat}
\preprint{}
\title{Controlling thermal expansion: A Metal--Organic Frameworks route}
\author{Salvador \surname{R. G. Balestra}}
\affiliation{Departament of Physical, Chemical and natural Systems, Universidad Pablo de Olavide, Ctra. Utrera km 1, 41013 Seville, Spain}
\author{Rocio \surname{Bueno--Perez}}
\affiliation{Departament of Physical, Chemical and natural Systems, Universidad Pablo de Olavide, Ctra. Utrera km 1, 41013 Seville, Spain}
\author{Said \surname{Hamad}}
\affiliation{Departament of Physical, Chemical and natural Systems, Universidad Pablo de Olavide, Ctra. Utrera km 1, 41013 Seville, Spain}
\author{David \surname{Dubbeldam}}
\affiliation{Van't Hoff Institute for Molecular Sciences, University of Amsterdam, Science Park 904, 1098 XH Amsterdam, }
\author{A. Rabdel \surname{Ruiz-Salvador}$^{~1,*}$}
\author{Sof\'ia \surname{Calero}$^{~1,}$}
\homepage[(ARRS) \url{rruisal@upo.es} and (SC) \url{scalero@upo.es}\\Visit: ]{http://www.upo.es/raspa/}
\date{12th October 2016}
\begin{abstract}
Controlling thermal expansion is an important, not yet resolved, and challenging problem in materials research. A conceptual design is introduced here for the first time, for the use of MOFs as platforms for controlling thermal expansion devices that can operate in the negative, zero and positive expansion regimes. A detailed computer simulation study, based on molecular dynamics, is presented to support the targeted application. MOF-5 has been selected as model material along with three molecules of similar size and known differences in terms of the nature of host--guest interactions. It has been shown that adsorbate molecules can control, in a colligative way, the thermal expansion of the solid, so that changing the adsorbate molecules induces the solid to have positive, zero or negative thermal expansion. We analyze in-depth the distortion mechanisms, beyond the ligand metal junction to cover the ligand distortions, and the energetic and entropic effect on the thermo-structural behavior. We provide an unprecedented atomistic insight on the effect of adsorbates on the thermal expansion of MOFs, as a basic tool towards controlling the thermal expansion.

Graphical TOC Entry:
\begin{figure}[h!]
	\begin{center}
	   \centering
	   \includegraphics[height=4.76cm]{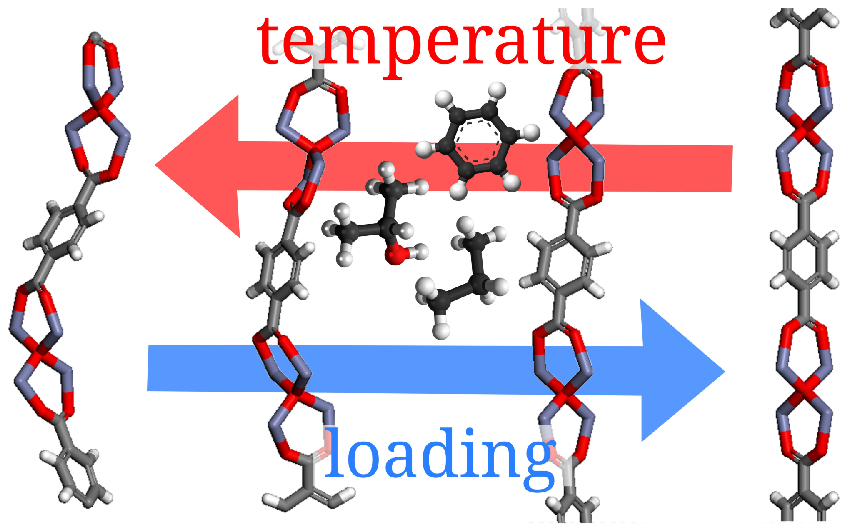}
	\end{center}
\end{figure}
\end{abstract}
\maketitle
\section{Introduction}
Thermal expansion (TE) mismatch is at the core of common mechanical failures in a wide range of systems \cite{han1996effect,darolia2013thermal,della2013copper,hu2011self}. This is an intriguing problem that needs to be solved for the development of new applications in advanced industries, such as aerospace and
microelectronics. To overcome this problem, it is desirable to have materials with controlled thermal expansion (CTE), a topic
that has been of high interest over decades \cite{han1996effect,della2013copper,kelly2006composite,tallentire2013systematic}. Currently, the most frequently used choice is composite materials,  whose TE can be controlled by
adjusting the amount of components having positive (PTE) and negative thermal expansion (NTE), in such a way that the overall behavior
fits the targeted necessities \cite{kelly2006composite}. Composite materials, however, have a high concentration of interfaces, which are weak
points, not only from a mechanical but also from a chemical point of view, mainly at medium and high temperatures \cite{banks201450th,sethi2015environmental}. There
is therefore a great interest in finding systems where the control of the thermal expansion can be achieved, without resorting to composite materials.

Recently, \citeauthor{tallentire2013systematic} successfully prepared cubic Zr$_{1-x}$Sn$_x$Mo$_2$O$_8$ solids, and found an unprecedented level of chemical control of the thermal expansion (negative, zero and positive) in a single phase, over a very wide range of temperatures \cite{tallentire2013systematic}. They started from the known fact that oxide frameworks, such as cubic \ce{ZrW2O8}, show NTE \cite{mary1996negative}, while the analogous \ce{SnMo2O8} is unique in showing PTE. Shortly after, \citeauthor{carey2014chemical} studied the TE of dehydrated \ce{Li-}, \ce{Na-}, \ce{K-}, \ce{Rb-}, \ce{Cs-} and \ce{Ag-} exchanged zeolite A and their purely siliceous analogue ITQ-29. \cite{carey2014chemical} They found that, in dehydrated state, NTE dominates in most cases, except for Li- and Ag-forms, which show a small PTE. On the other hand, PTE is observed in the presence of water. These approaches stimulate the search for materials that can be easily prepared and readily tuned to obtain ad hoc thermal expansion coefficients. Since a number of Metal-Organic Frameworks (MOFs) are known to show NTE \cite{carey2014chemical,collings2014geometric,lock2012effect,peterson2010local}, it is interesting to investigate whether MOFs can be used to develop systems with controlled thermal expansion. This is also stimulated by the recent study of \citeauthor{yot2014metal} that shows that MOFs can be used as a shock absorber \cite{yot2014metal}. In this context, exploiting the mechanical properties of MOFs emerges as a fascinating new branch in MOFs applications horizon.\cite{li2014research}

There are theoretical and experimental studies that show that MOFs exhibit interesting TE features. \citeauthor{dubbeldam2007exceptional} predicted exceptional NTE in MOF-5 using molecular simulations \cite{dubbeldam2007exceptional}, achieving a good agreement with the experimental behavior of the material loaded with \ce{CO2}, \ce{N2}, and \ce{Ar} \cite{rowsell2005gas}, results that were later validated experimentally for the empty framework \cite{zhou2008origin}. \citeauthor{yang2009crystallographic} found a MOF that undergoes PTE when desolvated, but it contracts in the presence of \ce{N2} at temperatures below 119~K, while it expands at higher temperatures \cite{yang2009crystallographic}. \citeauthor{joo2013volume} predicted a cell volume contraction due to van der Waals interactions of guest \ce{H2} molecules in MOF-5 \cite{joo2013volume}. \citeauthor{lock2012effect} found experimentally that the NTE observed in MOF-5 decreases with the amount of loaded helium \cite{lock2012effect}.

In addition, cell contraction in the breathing MOF MIL-53, induced by attractive host--guest interactions, has been identified \cite{boutin2010behavior,neimark2011structural}. And \citeauthor{grobler2013tunable}\cite{grobler2013tunable} observed that the extent of positive thermal expansion in an anisotropic MOF can be tuned by the adsorption of molecules, and that the size of the molecules is correlated with the induced change of the CTEs. All these findings suggest the hypothesis that MOFs can be tuned to be materials with CTE coefficients. In this study, we show at a theoretical level that controlling the nature and amount of adsorbed molecules in a MOF, the TE coefficient can be finely tuned to desirable behavior (PTE, NET or ZET, i.e. Zero TE). Note that large deformable MOFs, like MIL-53 \cite{serre2002very} or soft coordination polymer \cite{matsuda2004guest}, are not likely to be relevant solids for CTE, as their volume changes usually largely exceed the required adjustment of the TE-induced size mismatch.

As a model material we have selected MOF-5\cite{li1999design}, considering that not only it is an archetypal MOF, but most importantly, that molecular simulation methods perform very well in modelling its thermal behavior, including the NTE \cite{dubbeldam2007exceptional}. The adsorption of polar molecules, in MOF-5 \cite{bellarosa2012mechanism}, is known to occur in the proximity of the metal oxocluster. Since the formation of water clusters in this material provokes chemical instability \cite{bellarosa2012mechanism}, we have chosen an alcohol molecule, isopropanol (IPA), to study the influence of polar adsorbate--metal oxocluster interactions in its thermal behavior. Similarly, but for comparative purposes, we have also studied benzene (BEN) to observe the behavior associated to non-polar adsorbate--ligand interactions, as the role of the ligands in adsorption in MOF-5 has been identified to be of the same order than this of the oxocluster \cite{sarmiento2012surprising}.  Experimental results have shown that the ligands in MOF-5 also act as adsorption sites \cite{rowsell2005gas}. Other authors, using molecular simulation, found similar conclusions with regards to the presence of an adsorption site next to the ligands \cite{ray2012van,guo2010molecular}. A large discussion of the effect of metals and ligands on the adsorption properties of MOFs can be found in a recent review by \citeauthor{andirova2016effect} \cite{andirova2016effect}. In the present case, since benzene as adsorbate can interact with the benzene ring of the ligand via $\pi$--$\pi$ interactions\cite{he2015phenyl}, it is interesting to consider also a linear alkane, propane (PRO), as non-polar adsorbate with no particular interaction with the MOF. The overall behavior of MOF-5 when guest molecules are adsorbed is expected to be the result of a balance between three factors, namely the attractive host--guest interactions, the vibrational modes of the MOF structure (taking into account that they are likely to be affected by the adsorbates), and the repulsive contributions from adsorbate--framework collisions, which are more relevant at higher temperatures. We noted that MOF-5 collapses at relatively low external pressures\cite{hu2010amorphization}, which precludes its use at high mechanical stresses. Nevertheless, this system has other advantages. the material is an appropriate model system, considering the available experimental and theoretical literature on its thermal behavior, and on the other hand it can be used for controlling TE in small devices, such as those required in microelectronics.

\section{Conceptual Design, Methods and Computational Details}
In this work, we introduce a new concept in MOFs applications: their use as materials for controlling thermal expansion, in the three regimes, namely negative, zero and positive. As described in the introduction, there is a large amount of published studies accounting for interesting thermo--structural behaviors of MOFs, including changes of the thermal expansion with adsorbed molecules, which support the devised application. Our approach consists in the design of coatings made by MOFs, which are loaded with a certain amount of adsorbate molecules. These MOFs can have a desired, specific thermal expansion, covering all the range of behaviors, i.e. negative, zero and positive expansions. It is well known that a number of MOFs suffer damage upon desolvation, even leading to loss of crystallinity and structural collapse or amorphization \cite{ma2009freeze}. The presence of guest molecules in a MOF can enhance its mechanical resistance against framework collapse \cite{bennett2015mechanical}. Therefore, a careful selection of the MOF and the adsorbate molecules is essential. Nowadays there are several robust MOFs that have been proven to withstand cycles of solvation and desolvation, such as those studied by \citeauthor{khutia2013water} on MIL-101 \cite{khutia2013water}, and \citeauthor{begum2015water} \cite{begum2015water}. Chemical stabilization of certain MOFs, which are known to collapse otherwise, can be achieved, by applying solvent exchange to remove the pristine molecules in the pores resulting from the synthetic procedure. Such molecules might exert large capillary forces on desolvation, leading to collapse, but an exchange with weak-interacting solvents can prevent it \cite{mondloch2014zr}. We have paid attention to the known fact that the equilibrium loading of adsorbate molecules in a porous material depends on temperature, pressure, and the nature of the molecules. In this regard, since the planned application involves variations of external temperature, and implicitly also of external pressure, the device used for controlling thermal expansion must be operated without molecular exchange with the environment.

Simulations were performed with the RASPA code \cite{dubbeldam2016raspa}. The isosteric heats of adsorption of the guest molecules were computed after 500,000 sampling steps using the Widom Insertion Particle Method \cite{widom1963some}. Adsorbate--adsorbent interactions were modeled with Lennard--Jones (LJ) pairwise interatomic potentials, plus coulombic interactions. The values of the LJ parameters were calculated through Lorentz--Berthelot mixing rules, for which the force field parameters of the atom of the MOF were taken from UFF force field \cite{rappe1992uff}, and those of isopropanol, propane, and benzene molecules were taken from the OPLS-aa force field \cite{jorgensen1988opls,jorgensen1996development}. The LJ interactions were computed in the real space within a cut-off of 12~\AA, while the coulomb interactions were handled using the Ewald summation method \cite{ewald1921berechnung,tosi1964cohesion}. The atomic charges used for the molecules are the assigned by the selected force field, and the atomic charges for the MOF were taken from \citeauthor{dubbeldam2007exceptional} \cite{dubbeldam2007exceptional}.

The overall thermostructural behavior is expected to depend, particularly at high adsorbate loadings, on the adsorbate--adsorbate interactions, as well as on the framework properties and framework--adsorbate interactions. Therefore, we did not only pay careful attention to the force field used for the framework but also th that of the adsorbates. The force fields employed to model guest--guest interactions have been proven to model accurately the liquid phase of the compounds \cite{jorgensen1993monte,nemkevich2010molecular,caleman2011force}, so that we can rely on its validity modelling the dense phases within the pores.

Monte Carlo (MC) simulations in the Canonical (NVT) ensemble were conducted initially with one adsorbate molecule, in order to compute average occupational density profiles of each adsorbate. This was used as a tool for localizing the adsorption sites, and to calculate the binding energies of the adsorbates on the preferential sites. The maximum loading capacities were extracted from the saturation of the adsorption isotherms computed in the Gran Canonical ($\mu$VT) ensemble. To insert successfully the molecules in the system, the Configurational Bias Monte Carlo (CBMC) technique was used \cite{frenkel1996understanding}. NVT MC simulations were conducted with 20, 40, 60 and 80 \% of saturation of adsorbate molecules as starting configurations for the subsequent molecular dynamics simulations. MC simulations were run with 80,000 and 1,000,000 equilibration and production steps, respectively.

Molecular Dynamics (MD) simulations were performed in the Isothermal-Isobaric ensemble, (NPT), with isotropic cell fluctuations, using the fully flexible force field reported by \citeauthor{dubbeldam2007exceptional} \cite{dubbeldam2007exceptional} for the description of the thermo-structural properties of MOF-5 with and without adsorbate molecules. The integration of the equations of motion that generate the NPT ensemble was performed following the scheme of \citeauthor{martynatobiasklein} \cite{martynatobiasklein,martyna} A short time step of 0.5 fs was used to avoid the generation of abnormally large interatomic forces that might eventually induce partial collapse of the framework. A total of 200,000 steps (100 ps) were used for the equilibration of the systems and 10,000,000 steps (5 ns) for the production run. The adsorbate--adsorbent interactions were computed in the same way as used for the computation of the heats of adsorption. The structural data, acquired from the MD simulations, were analyzed with a home-made code, explicitly written for this purpose, which allows partitioning the cell length deformations among the different geometrical units composing the material.

\section{Results and Discussion}
\begin{figure}[h!]
	\begin{center}
	   \centering
	   \includegraphics[width=0.40\textwidth]{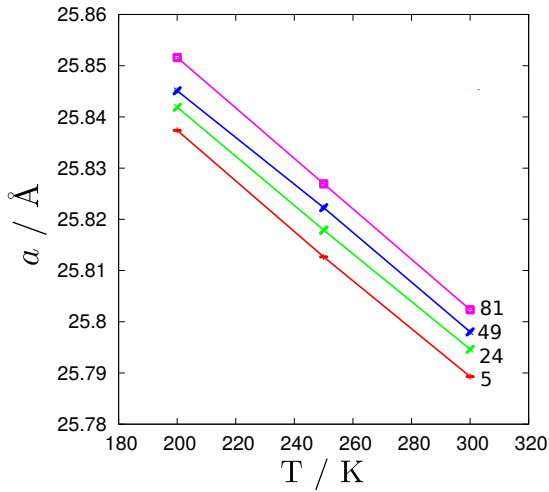}
	   \caption{\label{fig:helium}
	   Variation of the cell parameter, $a$, of MOF-5 with temperature, $T$. The MOF was loaded with 5, 24, 49 and 81 helium molecules per unit cell (red, green, blue and pink lines, respectively).
	   }
	\end{center}
\end{figure}

We first present some introductory results that are useful on the one hand for validating the theoretical methods, and on the other hand to describe the host--guest interactions between MOF-5 and the selected guest molecules. The closest available experimental results with which we can compare, regarding the variation of the thermal expansion of MOF-5 produced by changes in the amount of adsorbed molecules, are those reported by \citeauthor{lock2012effect} \cite{lock2012effect}. They employed helium flows at different rates to vary the amount of adsorbate molecules. We have therefore simulated this system to show the reliability of our computer simulation protocol, though it is worth mentioning that only qualitative comparisons can be established, as an accurate, quantitative estimation of the resident helium atoms inside the MOF in gas-flow operando experimental conditions is not possible. To obtain the best possible estimation, we calculate the number of helium atoms in the GCMC simulations at the pressures used to set the helium fluxes in the experiments of \citeauthor{lock2012effect} \cite{lock2012effect}. In Figure \ref{fig:helium} we plot the dependence of the cell volume of MOF-5 with temperature, for different amounts of helium atoms, (below 10\% of saturation capacity). It shows reasonable agreement with the experimental results \cite{lock2012effect}, since the correct NTE behavior is present, while there is an increase of cell parameters as the number adsorbed helium molecules increases. We did not get a decreasing behavior of the NTE coefficient as the number of guest molecules increases, since the number of flowing helium atoms per unit cell in the experiments is likely to be much higher than that we obtain for the equilibrium calculation (ca. 10\% of saturation) through GCMC. Below, we show that by increasing the number of adsorbate molecules, the NTE coefficient can indeed be tuned.

\begin{table*}[t]
\caption{\label{tab:energy_entropy_balance}
Calculated isosteric heats of adsorption, $\triangle H$, adsorption entropies, $\triangle S$, binding energies, $\triangle U$, binding energies at minimum energy configuration, $\triangle U_{\text{min}}$, and Henry coefficients, $K_{H}$ of benzene, isopropanol, and propane in MOF-5, at 300 and 500~K.
}
\begin{tabular*}{\textwidth}{@{\extracolsep{\fill}}ccccccc}
\\
 & $T$ / K & $\triangle H$ / & $\triangle U$ / & $\triangle U_{\text{min}}$ / & $T\triangle S$ /  & $K_{H}$ /                \\
 &         & kJ mol$^{-1}$   & kJ mol$^{-1}$   & kJ mol$^{-1}$                & kJ mol$^{-1}$     & mol kg$^{-1}$ Pa$^{-1}$  \\ \hline
Benzene     & $300.0$ & $-32.34$ & $-29.85$ & $-40.80$ & $-10.87$ & $1.381\times 10^{-3}$ \\
            & $500.0$ & $-28.16$ & $-24.0$  & -        & $-10.35$ & $1.082\times 10^{-5}$ \\ \hline
Isopropanol & $300.0$ & $-27.16$ & $-24.66$ & $-41.84$ & $-10.26$ & $2.174\times 10^{-4}$ \\ 
            & $500.0$ & $-22.52$ & $-18.36$ & -        & $-8.66$  & $4.18\times 10^{-6}$ \\ \hline
Propane     & $300.0$ & $-21.41$ & $-8.91$  & $-31.11$ & $-7.45$ & $6.7\times 10^{-5}$ \\
            & $500.0$ & $-19.46$ & $-15.3$  & - & $-7.59$ & $2.587\times 10^{-6}$ \\ 
\end{tabular*}
\end{table*}

\begin{figure}[h!tbp]
	\begin{center}
	   \centering
	   \includegraphics[width=0.48\textwidth]{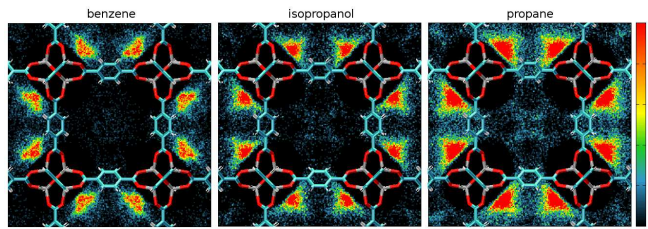}
	   \caption{\label{fig:colorin}
	   Average occupational density profiles of the center of mass of adsorbate molecules (benzene, isopropanol, and propane), in an $xy$--view. The cages located at the center and the corners of each snapshot correspond to the small cage. The framework atoms are superimposed to get a better understanding of the density profiles. The color code of the framework atoms is: carbon, blue; oxygen, red; hydrogen, white; zinc, grey respectively.
	   }
	\end{center}
\end{figure}

Using the selected probe molecules, we will provide a rationalization on the behavior found in Figure \ref{fig:helium}. As stated above, the nature and strength of the interactions of the guest molecules with the material are expected to play an important role in its overall thermo--structural behavior. Accordingly, MC simulations were used to study the adsorbate--adsorbent interactions for the selected three molecules. The heats of adsorption, adsorption entropies, Henry coefficients, and binding energies are reported in Table \ref{tab:energy_entropy_balance}. The strength of the host-guest interactions, as expected, increases with the number of non-H atoms. In general, single adsorbate molecules are preferentially located near the zinc atom of the oxocluster, in a corner also delimited by the adjacent atoms of the three benzene rings (Figure \ref{fig:colorin}). In the MOF-5 structure there are two types of cages, big and small, with different degreees of rotation of the benzene rings in the linker. Our results shows that the three guest molecules studied occupy mainly the big cages in first place. The distribution of molecules around the oxocluster is different between benzene, which spreads up to the benzene rings in the linkers, and isopropanol and propane, which are rather concentrated around the oxocluster. Likewise, the different nature the adsorption of isopropanol and propane is shown in the wider area that the latter occupies around the oxocluster. This is also supported by the different behavior found in the adsorption isotherms (Figure S2).

Along with the density profiles, the heats of adsorption and binding energies reveal that dispersive van der Waals interactions represent an important contribution to host-guest interactions. It is also noticeable that the stronger electrostatic interaction of isopropanol increases its binding energy. In the case of benzene, the $\pi$--$\pi$ interactions are responsible for the high binding energy. Nevertheless, the large porosity and heterogeneity of binding sites of this material lead to a much lower average interaction strength. This is also reflected in the entropy and Henry coefficients and from a structural point of view, this is revealed by slight changes of the atomic density near the oxocluster: isopropanol is more concentrated than propane at the corner of the oxocluster and benzene is also likely to occupy and intermediate position between the oxocluster and the benzene ring in the linker. We noted that at high loading conditions the sites in the small cages are also are occupied by guest molecules (Figure S1).
\begin{figure*}[h!tpb]
	\begin{center}
	   \centering
	   \includegraphics[width=0.8\textwidth]{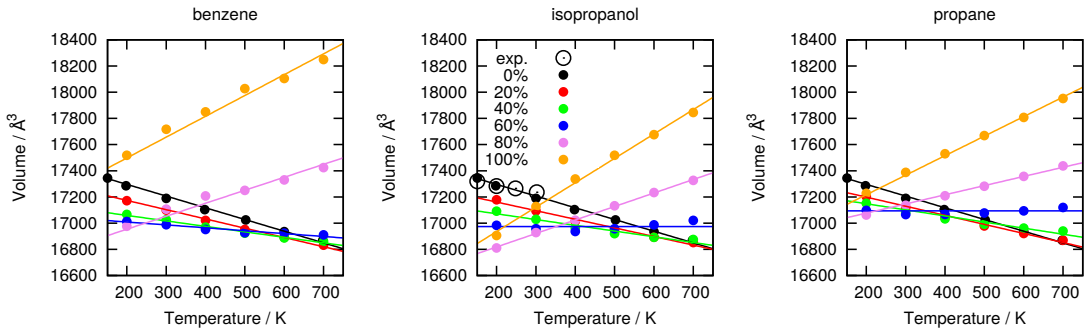}
	   \caption{\label{fig:TE}
       Cell volume as a function of temperature, for several loadings (in \% of saturation) of benzene, isopropanol, and propane. Experimental data are taken from references\cite{li1999design,eddaoudi2002systematic}. Solid lines represent linear regressions ( $r^2 > 0.9$ ). The color code is the same for the three figures.
	   }
	\end{center}
\end{figure*}
\begin{figure*}[h!tbp]
	\begin{center}
	   \centering
	   \includegraphics[width=0.85\textwidth]{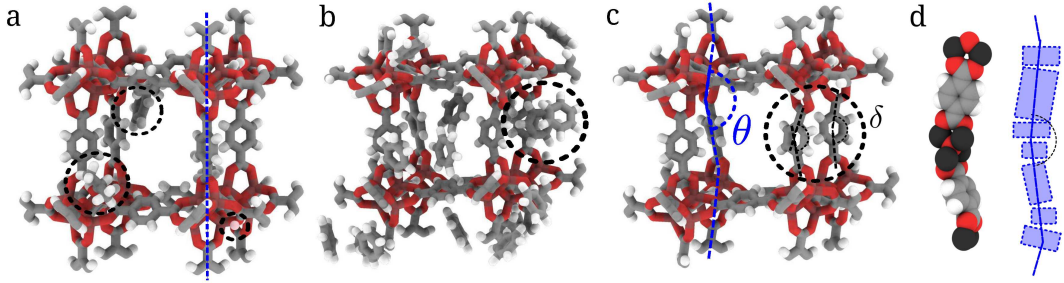}
	   \caption{\label{fig:strings}
       (\emph{a}) Structure of idealized undistorted MOF-5 with the guest molecules in their corresponding binding sites. (\emph{b}) Snapshot of the structure with 20\% of saturation of benzene. We show 
benzene molecules interacting with a benzene molecule of the linker via $\pi$--$\pi$ interactions. (\emph{c}) Distorted structure. Dashed blue lines represent a schematic view of the string of atoms crossing the cell, with different degrees of distortions $\theta$, and $\delta$ represents small distortions in the BDC linker. (\emph{d}, left) String of atoms crossing the cell. (\emph{d}, right) Schematic view of the line distorted string.
	   }
	\end{center}
\end{figure*}

The adsorption saturation capacities in MOF-5, as calculated by GCMC at high pressure, are 86, 98 and 105 molecules per unit cell, for benzene, isopropanol and propane, respectively.
As expected when we selected the molecules with similar molecular sizes, differences in the number of molecules do not exceed 20\%. Note that a cell volume change of the MOF should lead to a variation in the number of molecules in saturation conditions.
However, the variation of volume is tiny, compared with the total volume, and consequently the largest deviation in number of molecules is less than 3.2\%. 
We note that the comparison of the computed saturation capacity with experiments is only possible with benzene \cite{shim2012adsorption,eddaoudi2002systematic},as for the other two molecules there are not available published data.
The difference between the two experimental values available differ by 12.66\% (8.95 mol~kg$^{-1}$ and 10.24 mol~kg$^{-1}$, \citeauthor{shim2012adsorption} and \citeauthor{eddaoudi2002systematic}, respectively), which is no surprising considering that defects and residual dimethylformamide from the synthesis might alter the otherwise ideal adsorption capacity. This is also relevant that both experiments are conducted only to relatively low pressure values (less than 10 kPa). Our simulations are conducted in defect--free, completely desolvated frameworks, and the computed loading value at the pressure where adsorption saturation is observed (ca. 1~kPa) is within 15\% (11.96 mol~kg$^{-1}$) of the experimental data, which is a reasonable good agreement. In addition, as an important point for practical applications, it is useful to see that 98\% of adsorption is reached at experimentally accessible pressures (Figure S2).

Once the basic introductory data has been presented, we will carry out the analysis of the thermal expansion. The wide range of thermal expansion behaviors of MOF-5 loaded with benzene (BEN-MOF-5), isopropanol (IPA-MOF-5) and propane (PRO-MOF-5) is displayed in Figure \ref{fig:TE}. MOF-5 has a cubic space group and therefore we only plot the cell volume.
Three regimes are clearly observed: NTE, PTE, and ZTE. We found that the simulated thermal behavior of the bare framework is in fair agreement with experimental data.
We found that the simulated thermal behavior of the bare framework is in fair agreement with the experimental data.
This was also observed in a previous work \cite{dubbeldam2007exceptional}. It is worth noting that while the variation of the cell parameter of the mixed oxide solution of \citeauthor{tallentire2013systematic} is of the order of 0.5\%, in this system we find a larger variation, of 2.7\%, i.e. a range of variation five times larger.

A remarkable conclusion that can be drawn from Figure 3 is that the system shows a colligative behavior, as the qualitative behavior of the material does not depend on the choice of the adsorbed molecule. This is somehow surprising, as we noted that the interaction energies of the molecules with the framework as single entities (Table \ref{tab:energy_entropy_balance}) are different. Nevertheless, it is clear that the nature of the molecules allows a fine tuning of the volume variation, but the overall behavior is qualitatively the same for the three types of adsorbed molecules. The appearance of colligative behavior has not yet been reported in MOFs or coordination polymers. 
The key point in controlling the thermal dependence of the cell volume is the degree of guest loading. As shown in Figure \ref{fig:TE}, below 40\% the NTE regime is observed, and PTE appears above 80\%, while ZTE appears between these two values. In order to rationalize this interesting behavior it is instructive to focus firstly at very low temperatures. In this situation, the influence of the attractive host--guest interactions on the structure contraction is large. This is evidenced by cell volume values below saturation at 200 K that are lower than that of the bare structure. For instance, benzene molecules are attracted by the aromatic rings of the linkers, as shown in the snapshot of MOF-5 at 300~K, and a benzene loading of 20\% (Figure \ref{fig:strings}.b). In a second stage, it is interesting to pay attention to the behavior at loadings close to saturation, where the guest-induced molecular pressure on the framework is dominant, leading to cell volume increases, and the display of the PTE regime. As anticipated above, it is apparent that two effects compete in directing the thermal behavior: host--guest attraction and guest--induced pressure. The calculated TE coefficients systematically increase with a rise of the number of adsorbate molecules present in the structure. For the three molecules, at loadings of 20\% of saturation, the TE coefficients are around $-40\times 10^{-6}$ K$^{-1}$, they are nearly zero at 60\% loading, and the saturation values are $80\times 10^{-6}$ K$^{-1}$ for BEN and PRO, and $107\times 10^{-6}$ K$^{-1}$ for IPA. Details are given in the Supporting Information (Figure S3 and Table S2).
\begin{figure*}[h!tbp]
	\begin{center}
	   \centering
	   \includegraphics[width=0.8\textwidth]{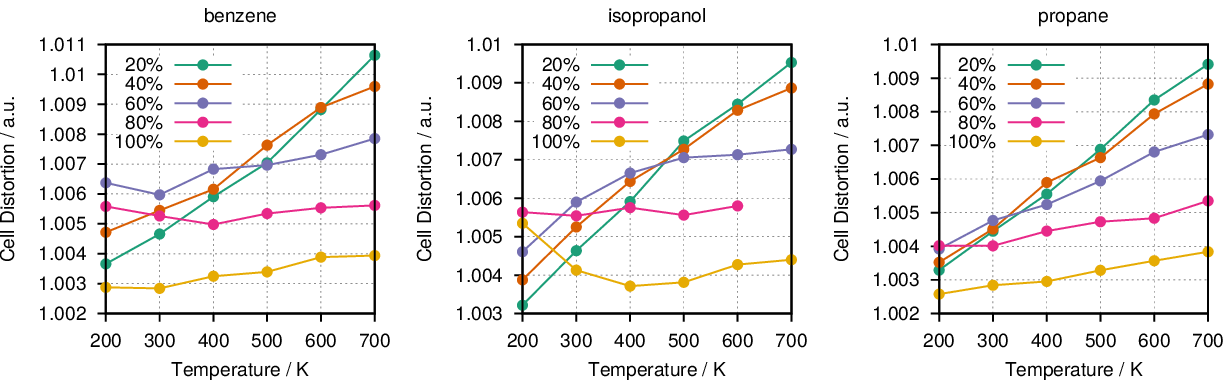}
	   \includegraphics[width=0.8\textwidth]{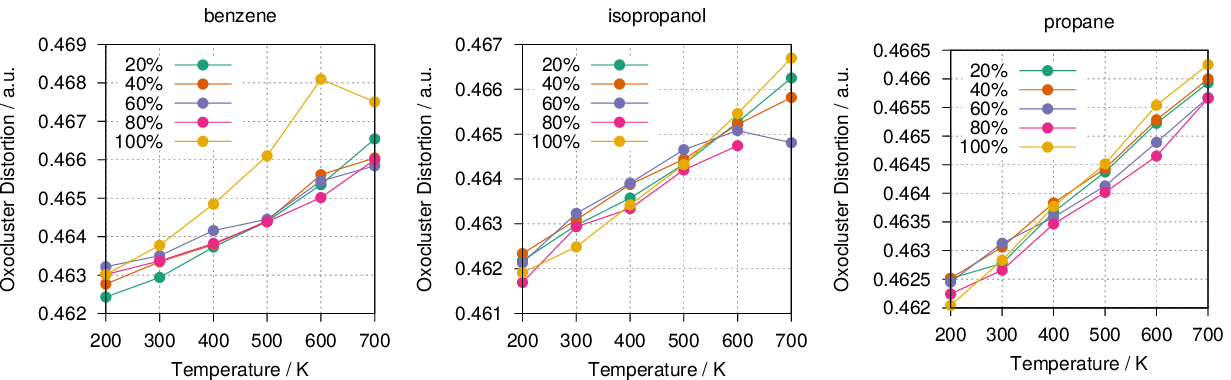}
	   \includegraphics[width=0.8\textwidth]{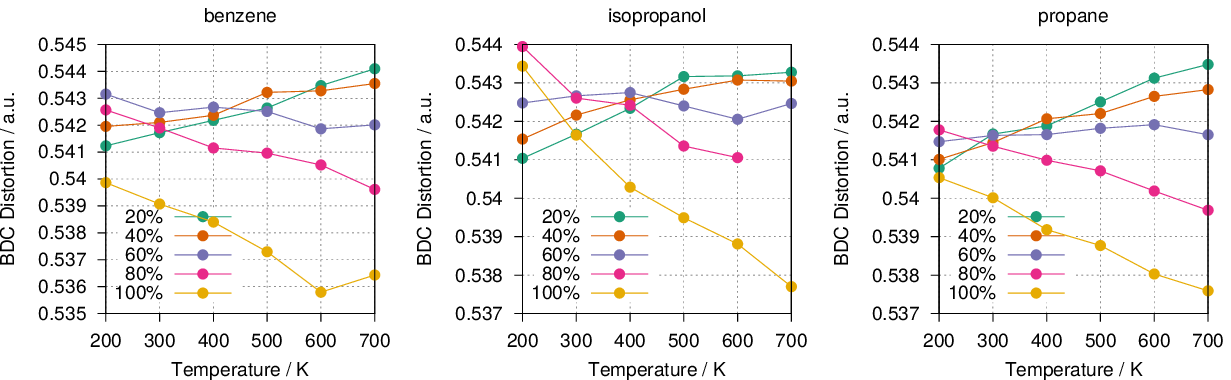}
	   \caption{\label{fig:distortions}
       Variation as a function of temperature of (top) the cell distortion, (middle) the oxocluster distortion, and (bottom) benzenedicarboxylate (BDC) distortion, for benzene, isopropanol and propane.    
       The distortions are defined in the text and in the ESI (Equation S1). Note than oxocluster plus BDC distortion sum about 1.
	   }
	\end{center}
\end{figure*}

To obtain futher insights into the thermal behavior of the system we analyze the relationship between molecular interactions, local deformations, and TE. From previous work, it is known that the NTE observed in MOF-5 is not associated with a concerted rotation of the oxocluster \cite{dubbeldam2007exceptional}, as it would be expected from the behavior in inorganic porous materials \cite{welche1998negative,pryde1996origin,barrera2005negative}. Although several studies have addressed this point \cite{lock2012effect,zhou2008origin,lock2013scrutinizing,lock2010elucidating,fang2014common}, the question remains still open, particularly when guest molecules are present. Here we perform a real space analysis of the system in order to get a better picture of the effect of the adsorbates over the structural behavior. As expected, in all cases the bonds of the solid elongate as temperature increases (Figure S5). Therefore a rationalization of the peculiar TE should be provided by the analysis of the collective behavior.

Using lattice dynamics DFT calculations \citeauthor{zhou2008origin}, found that a number of low energy modes are associated with deformation of BDC ligands \cite{zhou2008origin}. They did not identify them as a likely source of the observed NTE behavior, but they considered the BDC ligands as rigid units, as well as the oxocluster, and only linked the NTE with the junction flexibility. Further understanding was obtained by combining diffraction studies and cluster DFT calculations of the empty framework MOF-5 \cite{lock2010elucidating}. They suggested that the motions associated to low energy modes, which are responsible of NTE, are not only linked to the flexibility of connecting junctions, but also to intra-unit (ligand and oxocluster) motions \cite{lock2010elucidating}. Since our study comprises periodic MD simulations, which for structural dynamics studies is benefited by the absent of symmetry constrains within the unit cell \cite{balestra2015understanding}, here we focus on both the junction units and intra-unit deformations, as well as into their relation to the observed complex thermo--structural behavior. It is worth noting that this analysis can provide knowledge that will be useful for achieving control of the thermal expansion.

The structure of MOF-5 can be described as a 3-D grid structure, built up by strings that cross the cell, parallel to the axes (dashed blue line, in Figure \ref{fig:strings}.a). There is a degree of flexibility along the rods composing the strings, and it is easy to visualize that the further away the strings of atoms are from a perfect line (a schematic view in Figure \ref{fig:strings}.c,d), the shorter the corresponding cell axis will be. The rods represent the segments between the midpoints of the pair of oxygen atoms of each carboxylate group, and are depicted by blue bars in the schematic view of Figure \ref{fig:strings}.d. More details are given in the Supporting Information (Figure S4). 
In order to quantify the structural changes, we have plotted the ratio between the sum of the length of four rods and the length of the cell axis parallel to each line. We called this value "cell distortion", and it is displayed in Figure \ref{fig:distortions} (top) and in Equation S1. Two different regimes are found: an almost linear increase, at loadings below 80\%, and no variation, at higher loadings. The dominant role of the guest-induced pressure over the framework, at higher loadings, is evident from the figure, as the strings tend to expand as much as possible. Deviations from 1.0 are caused by thermal noise. At lower loadings, the collisions of these molecules with the framework are not enough to keep the strings extended. The most relevant effects are, instead, the flexibility of the rods junctions, and the internal deformations of the rods. The collisions play a key role in deforming the lines, as can be inferred by the rise of the line deformation parameter with an increase of temperature (Figure \ref{fig:strings}.c). The ZTE behavior (curves at ca. 60\% loadings) is then the result of the compensating effect between the increase of the atomic line deformation, i.e. the relatively smaller geometric line length, and the natural increase of the length of the bonds.

The central and bottom panels of Figure \ref{fig:distortions} show the relative size of the oxocluster and BDC, with respect to the cell size. The behavior of the oxocluster unit shows a monotonous variation with temperature, it is mostly independent of loading, for the three molecules. However, we observed that the effects of the loading on the shape and size of the BDC unit are radically different. At low loading, the relative distortion of the BDC unit increases with the increase of temperature and at high loadings, the relative distortion decreases. This is related to the ability of the BDC unit to be distorted. This observation is further supported by the analysis of the angle formed by the average points of the oxygen atoms in the carboxylic groups and the carbon atoms in the benzene ring (top panel Figure S7).

One important conclusion arisen from the structural analysis, in connection to the TE, is related to the flexibility of MOFs, particularly to the flexibility and deformations of both, the junctions between molecular units and the units themselves. There is an accepted understanding that flexibility in MOFs can be depicted in a mechanical view, based on the flexibility of the junctions connecting rigid units \cite{sarkisov2014flexibility}. However, this mechanical approach fails to explain the flexibility behavior for the isostructural MOFs MIL-47 and MIL-53, being the first rigid and the second flexible \cite{devautour2009estimation}. Our results indicate, moreover, that besides the primary source of flexibility associated to the units junctions, the nature of the units is also essential, and the deformations inside the units also contribute to the overall flexibility of the materials. A detailed description of the intra- and inter-units deformations in terms of relevant angles, distances and distortions parameters can be found in the Supporting Information Figures S5, S6, S7, and S8.

In our investigation of routes to control thermal expansion, we have shown that there is a clear connection between adsorbate loading, local and long range distortions, and thermo-structural behavior. We have shown that the material can be regarded as assembled by relatively rigid units, such as the oxoclusters and benzene rings, and by units with some degree of flexibility, such as the fragments composed by the acid groups with the connecting carbon atom from the benzene ring, and flexible junctions. Focusing on the strings of atoms (Figure \ref{fig:strings}.d), a large number of degrees of freedom can be associated to this particular structural motif, although their motions are constrained by the 3-D architecture of the network. In addition, the internal motion of each grid is restricted by the presence of atomic rings and clusters. The NTE thermal behavior of the empty framework can then be easily interpreted in terms of the Rigid Unit Modes (RUM) formalism  \cite{pryde1996origin,hammonds1997insights}, which supports the presence of cooperative modes. They generate disorder, making the atomic string to deviate more from the ideal line as temperature increases, as can be observed in Figure \ref{fig:distortions} bottom. In presence of adsorbates, host--guest attraction causes a degree of coordination between the movement of the adsorbate molecules and the flexible constituents of the material. This explains the observed behavior: At low loading the adsorbate molecules have large local mobility therefore increasing the NTE coefficients. And at high loading, the average position of the center of mass the adsorbate molecules is rather static due to the lack of available space, consequently ruling out the appearance of cooperative modes responsible of NTE. In the latter case, the close intermolecular distances provoke rocking motions that result in PTE.

On this basis, it is easy to rationalize why, at low loading, the qualitative behavior of the thermal expansion is similar for a wide range of temperatures. We observe that guest molecules are localized around certain positions, and have negligible impact on the host-host interactions that are associated to the NTE regime. Conversely, at high loading, there are "steric shielding effects" on the thermal expansion behavior, i.e. host--host interactions are seemingly weakened by the high number of guest molecules.
From a materials design point of view, we know that long range coulombic interactions can finely modulate the intrinsic TE properties of a MOF \cite{hamad2015atomic}. It is known that the charge distribution in a MOF can be tailored by changing the chemical composition \cite{kapelewski2014m2,gygi2016hydrogen,pham2014understanding,rosnes2015intriguing,pham2015dramatic}. For example, in a combined experimental and computational work, it was shown that varying the metal nature in \ce{M2}(m-dobdc) (\{M = \ce{Mg}, \ce{Mn}, \ce{Fe}, \ce{Co}, \ce{Ni}\} and  m-dobdc(4-) = 4,6--dioxido--1,3--benzenedicarboxylate) MOFs the polarity is systematically changed \cite{kapelewski2014m2}. On the other hand, we show here that despite the colligative behavior found for the thermo-structural properties, the nature of the molecules influences the change of the cell volume at given loadings.
Overall, we advance that exploiting the modulation of TE given by long-range, coulombic interactions, in connection with the guest-assisted control shown in this work, it is possible to achieve a large versatility in controlling the thermal expansion characteristics of the material.

\section{Conclusions}
In summary, we devised an approach with which it is possible to create systems with fine-tuned thermal expansion coefficients, thanks to the structural properties of Metal--Organic Frameworks. We found a complex interplay between competing effects, which permits the control of the thermal expansion. The attractive host--guest interactions induce the cell to contract, particularly at low loading. In these conditions, upon an increase of temperature, the thermal disorder increases and thus the coupled host--guest movements largely distort the atomic strings, which reduce the cell parameters, expressed as NTE. With increasing loading, the collisions of the guest molecules to the framework tend to direct the atomic strings to straight lines, and therefore inducing an increase of the cell parameter. At about 60\% of loading this effect is not enough yet to produce PTE, which, combined with the natural increase of bond distances, results in ZTE. At higher loadings, the increasing number of collisions leads to significant effects on the framework dynamics, much larger than the volume-reducing effect of the attractive host-guest interactions, so PTE is observed. In summary, we have found a surprising colligative behavior of the system, which determines the thermal expansion of MOF-5, featured by a general behavior that, for these similarly sizes molecules, is not particularly dependent on the nature of the adsorbed molecule.

\vspace*{1cm}
\hspace*{-0.45cm}
\textbf{Associated Content} \\
Supporting Information. Snapshots of MOF-5 saturated with adsorbed molecules, Figures and Tables with relevant geometrical and structural parameters. \\
\textbf{Author Contributions} \\
The manuscript was written through contributions of all authors. \\
\textbf{Funding Sources} \\
European Research Council, Spanish \emph{Ministerio de Economía y Competitividad} and Netherlands Council for Chemical Sciences.

\section{Acknowledgments}
The research leading to these results has received funding from the European Research Council under the European Union's Seventh Framework Programme (FP7/2007-2013) / ERC grant agreement nº [279520]), the Spanish \emph{Ministerio de Economía y Competitividad} (CTQ2013-48396-P), and the Netherlands Council for Chemical Sciences (NWO/CW) through a VIDI grant. SRG thanks the \emph{Ministerio de Economía y Competitividad} for his predoctoral fellowship.

\bibliography{biblio} 
\end{document}